\documentclass{elsart}
\input{psfig.sty}

\begin{document}
\begin{frontmatter}
\hyphenation{english}
\title{Probing Proton Strangeness with  
Time-Like Virtual Compton Scattering}
\author{Stephen R. Cotanch}
\address{Department of Physics, North Carolina State
University, Raleigh, NC  27695 USA}
\author{Robert A. Williams}
\address{  Nuclear  Physics Group, Hampton University, 
Hampton, VA 23668 USA \\ 
and \\
Jefferson Lab,
12000 Jefferson Avenue, Newport News, VA 23606 USA } 
\date{\today}
\maketitle
\begin{abstract}
We document that $p(\gamma,e^+e^-)p$ measurements will yield new,
important information about the off-shell time-like nucleon form factors,
especially in the $\phi$ meson region ($q^2 = M^2_{\phi}$) governing the   
$\phi N$ couplings $g^{V,T}_{\phi N N}$. 
Calculations for $p(\gamma,e^+e^-)p$, utilizing vector meson dominance,  predict
measurable $\phi$ enhancements at high $|t|$  compared to the expected $\phi$  
background production from 
$\pi$, $\eta$ and Pomeron exchange. The  
$\phi $ form factor contribution 
generates a novel experimental signature for 
OZI violation and the proton
strangeness  content.  The $\phi N$ couplings
are determined independently from a combined analysis of
the neutron electric form factor and recent high $|t|$
$\phi$ photoproduction.
The $\pi$, $\eta$ and Pomeron transition form factors are also
predicted and the observed $\pi$ and $\eta$ transition
moments are reproduced.    
\end{abstract}

\vspace{1.5cm}
\scriptsize{
 
}
\end{frontmatter}

\newpage




\hspace*{\parindent}

Compton scattering, an elegant time-honored topic, is an effective probe of
hadron structure. In this paper we calculate a striking effect 
for a less investigated version of this process,
time-like virtual Compton scattering [TVCS], $p(\gamma,e^+e^-)p$.  Invoking
vector meson dominance [VMD], we predict distinctive, dual peaked cross
section resonances for time-like virtual photon four-momentum 
spanning the
vector meson masses ($q^2 \sim M_{V}^2$ for $V=\rho , \omega , \phi$).
In our Quantum Hadrodynamic [QHD] model these narrow, order of magnitude enhancements
arise from vector mesons, associated with the outgoing virtual photon, coupling in the
$t$ channel to
$\pi$,
$\eta$ and Pomeron
$\mathcal P$ exchanges, and also to the proton through the time-like form factors
($F_1^p,F_2^p$ or
$G_E^p,G_M^p$) in the $s$ and $u$ channels. There is no nucleon form factor data for
$0 \leq q^2 \leq 4 M_{p}^2$ since all measurements to date have utilized the
two-body annihilation processes 
$e^+ e^- \leftrightarrow N \bar{N}$. However, $p(\gamma,e^+e^-)p$ 
involves a three-body final state with an essentially unrestricted 
virtual photon mass $q^2 \geq 4 M_{e}^2 \sim 0$.  The dramatic VMD resonant
signature was also noted in our previous analyses ~\cite{bwscprl,bwscnpa} of
$p(\pi^-,e^+e^-)n$ for experiments with hadron beams.  This
work extends that study and affirms 
TVCS measurements at electromagnetic facilities can provide 
similar information.
 
Knowledge of the nucleon time-like form-factors,
for both on and off-shell nucleons, permits a direct assessment of
the nucleon strangeness content.  Nucleon strangeness is quantified by 
the nucleon
matrtix elements, $\langle N|\bar{s} \Gamma_n s |N \rangle$, involving the
$n = 1,2, .. 16$ Lorentz
bilinear covariants.  Here we address the vector and tensor elements, $\gamma_\mu$ 
and $\sigma_{\mu \nu}$, which respectively corresponds to the coupling constants
$g^{V}_{\phi N N}$ and $g^{T}_{\phi N N}$, since the
$\phi$ is predominantly $s\bar {s}$.  These coupling constants completely
specify the QHD $\phi N N$ Lagrangian 
\begin{equation}
 \mathcal L_{\phi N N} = g^{V}_{\phi N N} \bar{N} \gamma _{\mu} N \phi^{\mu} +
 g^{T}_{\phi N N}  \bar{N} \sigma _{\mu \nu} N [\nabla
^{\mu}\phi^{\nu} -
\nabla ^{\nu}\phi^{\mu}]
\end{equation}

which we use for our cross section predictions presented below.  By
extracting these couplings from the nucleon form factors using VMD and from
non-diffractive $\phi$ photoproduction ~\cite{Ellis1,Henley,Titov}, one can quantify the
hidden strangeness in the nucleon.  Clearly if the probability of an $s\bar{s}$ pair in
the proton is zero or extremely small,
$\phi N$ coupling will be significantly
suppressed due to the dominant $s \bar{s}$ structure of the
$\phi$ and the OZI rule ~\cite{OZI} (i.e. suppressed interactions 
between hadrons with different quark flavors).  However, sizeable OZI violations
($\phi$ production/coupling assuming $\langle N|\bar{s} \Gamma_n s |N
\rangle$ = 0) have been observed in $p \bar{p}$ annihilation experiments at LEAR 
by the ASTERIX, Crystal Barrel and
OBELIX collaborations ~\cite{Asterix}.
These results are consistent with analyses of other independent
experiments such as the EMC deep inelastic $\mu p$ 
scattering~\cite{EMC}, elastic $\nu p$ 
scattering at BNL~\cite{Ahrens} and measurements of 
the $\pi N$ sigma term~\cite{Donoghue} which all
support an appreciable strangeness component in the proton.
Reference~\cite{Ellis2} reviews
the evidence for hidden strangeness. 

Theoretically, there have been several strangeness 
calculations ~\cite{Titov,GeigerIsgur,Leinweber,mm,sw,mh,Speth} of both 
${\phi N}$ couplings and nucleon properties (spin, radius, magnetic moment)
using a variety of approaches.  Many of these analyses have been formulated in
terms of  hyperon Fock components for the nucleon, 
$|KY \rangle (Y = \Lambda, \Sigma)$, and/or meson loop contributions using
dispersion relations.  While loop
cancellations are appreciable, theoretical uncertainty precludes a robost
quantitative assessment of strangeness and additional
study appears necessary.  We will also address this in a future communication using
a many-body, relativistic field theoretical  quark/gluon formalism that implements 
chiral symmetry ~\cite{flsc}.  

Here we adopt a phenomenological approach which directly determines the
${\phi N}$ couplings from data and then predicts $p(\gamma,\gamma_v)p$.
We first compute the hadron form factors utilizing a generalization 
~\cite{WKL,WPT} of the VMD 
model developed in
ref.~\cite{Gari}.
The $SU_{F}(3)$ symmetry relations and 
Sakurai's universality hypothesis are incorporated to describe the baryon
octet EM form factors. 
A good description of the data is obtained using 
the specific vector meson-nucleon couplings,
$C_{\rho}(N) = 0.4$, $C_{\omega}(N) = 0.2$ and $C_{\phi}(N) = -0.1$, 
where $C_{V}(N) = g_{VNN}/f_{V}$ is the ratio of the vector meson-nucleon hadronic 
coupling, $g_{VNN}$, to the meson-leptonic decay constant, $f_{V}$.
From the $\phi \rightarrow e^+ e ^-$ decay width 
we extract the $\phi$ decay constant, $f_{\phi} = -13.1$, 
yielding the $\phi N$ vector coupling, $g^V_{\phi N N} = 1.3$.  Because of its
relatively small size, the neutron electric form factor, $G_{E}^{n}$,
is very sensitive to $\phi N$ coupling. As Fig. 1 indicates,
there is considerable uncertainty to both the vector and tensor, 
$g^T_{{\phi}NN} = \frac {\kappa^T_{\phi}}{M_\phi} g^V_{{\phi}NN}$, coupling constants.
New, but still preliminary, $G_{E}^{n}$ data indicates an even smaller form
factor which suggests a larger $\phi N$ coupling (inverse relation).  Clearly,
a precision $G_{E}^{n}$ measurement will significantly constrain the
values of $g^V_{{\phi}NN}$ and tensor moment $\kappa^T_{\phi}$.  However, by
analyzing
recent $\phi$ photoproduction data (see below) we have reduced part of this uncertainty
leading to the values $g^V_{\phi N N} = 1.3$,
$g^T_{\phi N N} = 2.3$ and $\kappa^T_{\phi} = 1.8$.   
Our vector $\phi N$ coupling constant relative to 
$\omega N$ is $g^{2}_{\phi N N}/g^{2}_{\omega N N} = 0.14$, slightly
smaller but still consistent with ref.~\cite{Ellis1}.

We also use VMD for the meson transition form factors entering the 
$t$ channel cross section contributions.  While the proton's
strangeness content remains uncertain, the  $u$ and $d$ quark components of the
$\phi$ and $\eta$ are much better known (see below). 
The pseudoscalar, $\gamma \pi \rightarrow \gamma_v$, $\gamma \eta \rightarrow \gamma_v$ 
and Pomeron, $\gamma {\mathcal P} \rightarrow \gamma_v$ transition form factors 
are computed from $\rho$, $\omega$ and $\phi$ vector meson propagators with 
couplings determined directly from the 
$\phi \rightarrow \gamma \pi$, $\phi \rightarrow \gamma \eta$,
$\omega \rightarrow \gamma \pi$, $\omega \rightarrow \gamma \eta$,
$\rho \rightarrow \gamma \pi$ and $\rho \rightarrow \gamma \eta$
decay widths~\cite{PDG}.
The leptonic decays $\phi \rightarrow e^{+} e^{-}$ and
$\omega \rightarrow e^{+} e^{-}$ together with the photon radiative 
decays $\pi \rightarrow \gamma \gamma$ and $\eta \rightarrow \gamma \gamma$ 
provide a consistency check on the VMD 
$\pi$ and $\eta$ transition form factors because of the normalization 
conditions
\begin{equation} 
\kappa_{\pi \gamma \gamma} = \frac{\kappa_{\rho \pi \gamma}}{f_{\rho}} 
\;+\;   \frac{\kappa_{\omega \pi \gamma}}{f_{\omega}} 
 \;+\;   \frac{\kappa_{\phi \pi \gamma}}{f_{\phi}}\ , \;\;\;\;\;
\kappa_{\eta \gamma \gamma} = \frac{\kappa_{\rho \eta \gamma}}{f_{\rho}} 
\;+\;   \frac{\kappa_{\omega \eta \gamma}}{f_{\omega}} 
 \;+\;   \frac{\kappa_{\phi \eta \gamma}}{f_{\phi}} \ . 
 \end{equation}
Using the most recent data~\cite{PDG} VMD predicts
the  moments $ \kappa_{\pi \gamma \gamma} \;=\; 0.30$,  
$\kappa_{\eta \gamma \gamma} \;=\; 0.27$,
$ \kappa_{{\mathcal P} \gamma \gamma} \;=\; 0.11$  which are in excellent 
agreement with the observed values $\kappa_{\pi \gamma \gamma} \;=\; 0.27 $,
$\kappa_{\eta \gamma \gamma} \;=\; 0.26$ ($ \kappa_{{\mathcal P} \gamma \gamma}$
has not been measured). 
The VMD couplings are summarized in Table I.
The Pomeron radiative transition moments have been extracted from a $\phi$
photoproduction analysis~\cite{WilliamsPhi}.  


\noindent{Table I. }
Vector meson transition moments and decay constants.
\vspace{0.15in}
\begin{center}
\begin{tabular}{|c||c|c|c|c|}
\hline
$V$ & $\kappa_{_{V} \pi \gamma}$ & $\kappa_{_{V} \eta \gamma}$ & 
$\kappa_{_{V {\mathcal P}} \gamma}$ & $f_{_V}$  \\
\hline
$\rho$ & 0.901 & 1.470  & 0.62 & 5.0 \\
$\omega$ & 2.324 & 0.532 & 0.62 & 17.1 \\
$\phi$ & 0.138 & 0.715 & 0.62 & -13.1 \\
\hline
\end{tabular}
\end{center}

Figures 2 and 3 display our calculated
proton form factors along with available data. 
Note for $q^2 = 1.0 \; GeV^2 \cong M^2_\phi $, both $G_{E}^{p}$ and $G_{M}^{p}$ 
exhibit significant, narrow resonances.
In the OZI limit (zero $\phi$ coupling),
the form factors are essentially unchanged in the space-like region
and the time-like $\phi$ peaks vanish.  The absolute values are plotted since the
form factors are complex in the vector meson region. The neutron time-like form factor
behavior is similar  but, as discussed above,  space-like 
$G_{E}^{n}$  is very sensitive to $C_{\phi}(N)$.  

Since $\phi$ photoproduction and 
TVCS have the same quantum numbers,
$\gamma(q,\lambda)$ + $p(p,\sigma) \rightarrow V(q',\lambda')$ + 
$p(p',\sigma')$, $V = \phi $  or $  \gamma_v$, 
the $cm$ cross section for both  is
\begin{equation}
\frac{d\sigma}{dt} \;=\; \frac{ M_{p}^2 \;(\hbar c)^2 }{64 \pi \; s \; 
|{\bf q}|^2}  \;  
\sum_{\lambda' \lambda \sigma' \sigma} \;
|T_{\lambda' \sigma' \lambda \sigma}|^2 \; ,
\end{equation}
with $q$, $p$, and $\lambda$, $\sigma$, denote the corresponding 4-momenta and  
helicities, respectively.        The photoproduction helicity amplitude, 
$T_{\lambda' \sigma' \lambda \sigma}$, is
\begin{equation}
T_{\lambda' \sigma' \lambda \sigma} \equiv \epsilon_{\mu}(\lambda) \; 
\phi_{\nu}^{*}(\lambda') \; {\mathcal H}^{\mu \nu}_{\sigma' \sigma} \; , 
\end{equation}
where $\epsilon_{\mu}(\lambda)$ and $\phi^*_{\nu}(\lambda')$ 
are the initial photon and final virtual photon (or $\phi$) 
polarization 4-vectors in the helicity basis, respectively.
The hadronic current tensor, ${\mathcal H}^{\mu \nu}_{\sigma' \sigma}$, 
is evaluated at tree level from the $s$, $t$ and 
$u$ channel QHD diagrams. Working 
in the $cm$ system ({\bf q + p = q' + p' = 0}), 
with the z-axis along $\bf q$, 
the photon polarization vectors are 
\begin{equation}
\epsilon_{\mu}(\lambda) =
 - \frac{\lambda}{\sqrt{2}} (0,1,i\lambda,0)  \;\;\;\;\;\;  
 (\lambda = \pm) \;,\\
\end{equation}
\begin{equation}
 \phi^*_{\mu}(\lambda') = \frac{\lambda'}{\sqrt{2}}
  (0,-\cos \theta_{cm}, i \lambda', \sin \theta_{cm}) \;\;\;\;\;\; 
 (\lambda' = \pm) \; ,  \\
\end{equation}
\begin{equation}
\phi^*_{\mu}(\lambda') = \frac{1}{\sqrt{q'^2}}
({\bf |q'|},q'_0 \sin \theta_{cm},0,q'_0 \cos \theta_{cm}) 
\;\;\;\;\;\; (\lambda' = 0) \; .
\end{equation}

Evaluating the QHD Lagrangian diagrams
generates the following $s,t$ and $u$ channel contributions to 
the hadronic current tensor.

\underline{$t$ channel $0^+$ Pomeron (${\mathcal P}$) exchange:}
\begin{equation}
{\mathcal H}^{\mu \nu}_{\sigma' \sigma} =
\Gamma_{{\mathcal P}} \; \Pi_{{\mathcal P}}(t)
 \left(\frac{s-s_{th}}{s_{0}}\right)^{\alpha(t)} 
 \bar{u}(p',\sigma') u(p,\sigma) \; 
[\; q\cdot q' \; g_{\mu \nu} - q'_{\mu} q_{\nu} \;] \; .
\end{equation}  


\underline{$t$ channel $0^-$ meson ($x$ = $\pi^{0},\eta$) exchange:}
\begin{eqnarray}
{\mathcal H}^{\mu \nu}_{\sigma' \sigma} &=&
\frac{ \Gamma_x \; F_{t}(t;\lambda)}
{M_{\phi}[(p'-p)^2 - M_{x}^2]} \;
\bar{u}(p',\sigma') \; \gamma_{5} \; u(p,\sigma) \;
\epsilon^{\mu \alpha \nu \beta} q_{\alpha} q'_{\beta}  \; .
\end{eqnarray}

\underline{$s$ channel proton ($p$) propagation:} 
\begin{eqnarray}
{\mathcal H}^{\mu \nu}_{\sigma' \sigma} &=& 
\bar{u}(p',\sigma') \; [ \Gamma_1' \;\gamma^{\nu} \;+\; 
i \Gamma_2'  \sigma^{\nu \alpha} q'_{\alpha} \;] \; 
\nonumber \\
&& \times \; \frac{ (p+q)\cdot \gamma + M_{p}}{(p+q)^2 - M_{p}^2 + 
\Sigma_p(s)} 
\; [\; \Gamma_1 \gamma^{\mu} \;+\;  
i \Gamma_2 \sigma^{\mu \beta} q_{\beta} \;] \;
u(p,\sigma) \; .
\end{eqnarray}  

\underline{$u$ channel channel proton ($p$) propagation:} 
\begin{eqnarray}
{\mathcal H}^{\mu \nu}_{\sigma' \sigma} &=& 
\bar{u}(p',\sigma') \; [ \Gamma_1 \;\gamma^{\mu} \;+\; 
i \Gamma_2 \sigma^{\mu \beta} q_{\beta} \;] \; 
\nonumber \\
&& \times \; \frac{ (p'-q)\cdot \gamma + M_{p}}
{(p'-q)^2 - M_{p}^2 + \Sigma_p(u)} 
\; [\; \Gamma_1' \gamma^{\nu} \;+\;  
i \Gamma_2' \sigma^{\nu \alpha} q'_{\alpha} \;] \;
u(p,\sigma) \; .
\end{eqnarray}  
The scalar ($0^+$) Pomeron 
coupling to the ${\mathcal P}NN$, $\bar{u}(p',\sigma') u(p,\sigma)$, and 
$\phi {\mathcal P} \gamma$,
$F_{\mu \nu}^{\gamma} F_{\phi}^{\mu \nu}$, vertices is represented by
$\Gamma_{\mathcal P}$, which is a product of Pomeron hadronic and electromagnetic
coupling constants.
The energy dependence, $(s/s_{0})^{\alpha(t)}$, follows from Regge theory
~\cite{Regge} with Pomeron trajectory, 
$\alpha(t)$ = .999 + .27 $\  GeV^{-2} \ t$,
which reproduces established high energy 
diffractive data.    
To describe the less well known low energy dependence, we introduce
the parameter $s_{th}$ ($0 \leq s_{th} \leq s_{0}$).
Taking the reference energy $\sqrt{s_{0}}$ as the 
production threshold, $s_{0} = (M_{V} + M_{p})^2$ 
($M_V = \sqrt{q'^2}$ or $M_{\phi}$ for $V = \gamma_v$ or $\phi$), 
we find the $\phi$ photoproduction data clearly selects the maximum value, 
$s_{th} \rightarrow
s_{0}$.
The $\Gamma$ factors in Eqs. (8-11) are products of hadronic
and electromagnetic coupling constants appropriate for either 
$V = \phi, \gamma_v$ and are listed
in Table II.  


\noindent{Table II. Effective vertex couplings. 
$\Gamma_1 = e = \sqrt(4 \pi \alpha_e)$,  $\Gamma_2 = \frac{e \; \kappa_p}{2
M_p}$, $\kappa_{p} = 1.793$, $g_{_{\mathcal P} NN} = 44.0$, $g_{\pi NN}= 13.8$ and
$g_{\eta NN} = 7.5$.}
\small

\vspace{0.15in}
\begin{center}
\begin{tabular}{|c|c|c|c|c|c|c|}
\hline
$V$ & $\Gamma_{_{\mathcal P}}$ & $\Gamma_{\pi}$ & $\Gamma_{\eta}$ & 
 $\Gamma_1'$  
& $\Gamma_2'$  \\
\hline
$\phi$ 
& 
$g_{_{\mathcal P} NN} (\frac{ e \; \kappa_{\phi {\mathcal P}  \gamma}}{M_{\phi}}) $
& 
$g_{\pi NN} (\frac{ e \; \kappa_{\phi \pi \gamma}}{M_{\phi}}) $
&
$g_{\eta NN} (\frac{ e \; \kappa_{\phi \eta \gamma}}{M_{\phi}}) $ 
&
$g_{\phi N N}$
& 
$g_{\phi N N} (\frac{\kappa^T_{\phi}}{M_{\phi}}) $
\\
$\gamma_v$ 
& 
$g_{_{\mathcal P} NN} (\frac{ e^2 \; \kappa_{_{\mathcal P} \gamma \gamma}}{M_{\phi}}) $
& 
$g_{\pi NN} (\frac{ e^2  \; \kappa_{\pi \gamma \gamma}}{M_{\phi}}) $
&
$g_{\eta NN} (\frac{ e^2 \; \kappa_{\eta \gamma \gamma}}{M_{\phi}}) $
&
$e \; F_1^p(q'^2)$
& 
$(\frac{e \; \kappa_{p}}{2 M_{p}})  F_2^p(q'^2) $
\\
\hline
\end{tabular}
\end{center}

\normalsize

The effective Pomeron propagator, $\Pi_{{\mathcal P}}(t)$,
describes~\cite{WilliamsPhi} scalar exchange 
\begin{equation}
\Pi_{{\mathcal P}}(t) \;=\; \frac{e^{\beta t}}{t - M_{{\mathcal P}}^2} \; ,
\end{equation}
and reproduces the  known diffractive $t$
dependence  
using the lightest scalar glueball mass,  $M_{\mathcal P} = 1.7 \; GeV$, 
and Pomeron slope 
$\beta$ = .27 $GeV^{-2}$.   
The pseudoscalar $t$ channel form factor, $F_{t}(t;\lambda)$, governs hadronic 
structure  and is necessary 
for the correct  4-momentum transfer dependence 
in meson photoproduction~\cite{Li,Lu}. 
Covariance and crossing symmetry are preserved using  
\begin{equation}
F_{t}(t;\lambda) = \frac{ \lambda^4 + t_{min}^2 }{\lambda^4 + t^2} \; ,
\end{equation}
normalized to unity at 
$t_{min} = t \ (\theta^{c.m.}_{\gamma \phi} = 0)$.
From $p(\gamma,\phi)p$ data, the optimum cutoff parameter is 
$\lambda = 0.7$ GeV.
For the $s$ and $u$ channels we
describe the off-shell proton by a self-energy correction, $\Sigma_p$, to the propagator.
To maintain both gauge invariance and the correct on-shell proton mass, 
$\Sigma_p$ must vanish at the proton pole and also be an
odd function of $(s - M_p^2)$
\begin{equation}
\Sigma_p(s) \;=\; \alpha_{off} \;  
\frac{(s - M_p^2)^3}{M_p^4} \; . 
\end{equation}       
The dimensionless off-shell parameter, $\alpha_{off} = 1.29$, 
was also adjusted for optimal 
agreement with recent, and higher $|t|$, $\phi$ photoproduction data~\cite{hi-t-phi}.
shown in Fig. 4.  Note the sensitivity and relative contributions
from the $|t|$ channel Pomeron $\mathcal P$ (dense dotted curve), $\pi$ (short dashed
curve),
$\eta$ (spare dotted curve) and $s$ channel $\phi N$ coupling (long dashed curve). 
Since the
$t$ channel processes are suppressed at high $|t|$, the latest data
permits more stringent constraints on the $\phi N$ coupling,  yielding
the values quoted above.  It would be extremely useful to have
measurements for $|t|$ $>$ 4 $GeV^2$ where our model
predicts an enhancement  from
interference between the vector and tensor $\phi N $ couplings.

\newpage 

Figures 5 and 6 represent our key results and display the TVCS cross
sections versus final proton lab angle and initial photon lab energy,
respectively.  Notice in Fig. 5 the dual peak resonant signature
due to the quadratic relation between $q^2$ and recoil proton lab angle.
The smaller angle $\phi$ peak, corresponding to high $|t|$, is dominated by the
$u$ channel (proton propagator) $g_{\phi NN}$ coupling (dense dotted curve 
labeled $p$).  This is our prediction for measuring the proton
strangeness content.  The two other peaks near $25^o$ and $30^o$, 
have lower $|t|$ and respectively represent $\phi$ and $\omega$ coupling
to mesons in $t$
channel exchange.  These two peaks embody established results
and therefore constitute the expected $\phi$ production background.
Figure 6 demonstrates the cross section energy dependence at
maximum $|t|$ on the $\phi$ resonance ($q^2 = M^2_{\phi}$).  Our model clearly
predicts $u$ channel domination (curve labeled $p$)  above $2.5 \ GeV$ lab energy, 
enabling direct extraction of $g_{\phi NN}$.

We wish to stress that our $\phi$ dual peak signature and magnitude follows from
VMD only and is independent of the underlying dynamical model.  While
our results utilized QHD, we submit similar findings will occur
for other models.  
Assuming only the validity of VMD, we confidently predict that a measurement
of the high $|t|$ TVCS cross section ratio $R = \sigma (q^2 = M^2_{\phi})/\sigma (q^2 =
M^2_{\omega})$ would yield a result proportional to $g^2_{\phi NN}/g^2_{\omega NN}$.
We have confirmed this numerically in our model, $R = 0.14 \ f$ (where f is a
kinematic quantity of order unity).  This is over an order of magnitude
larger than the OZI (no nucleon strangeness) prediction \cite{Ellis2}, $R = tan^2 \ \delta
\ f = 0.0042 \ f$, 
where $\delta = 3.7^o$ is the deviation  from the ideal quark flavor mixing angle in the
$\phi$.  Thus, TVCS experiments are an excellent
means for probing the proton
strangeness content.

Note also that TVCS probes the (half) off-shell 
proton form factors which, depending on kinematics, can be
quite different from the on-shell form factors used in this
study.  Reference~\cite{korchin} addresses this issue further,
including ambiguity in off-shell formulation.  Here we simply
point out that both on and off-shell form factors contain
important information about $\phi N$ coupling in the time-like
region and that the above ratio $R$ will be less sensitive
than the form factors to off-shell effects.  Also, by comparing
the TVCS off-shell form factors to the known on-shell $G^p_E$, $G^p_M$
for $q^2 \ge 4 M^2_p$, off-shell effects can be directly assessed.
A complete analysis also requires  $N^*$  resonances.  However,
by choosing large $s$ and $|t|$  kinematics  with $u \approx 0$,
only the lightest baryon resonances will compete with the proton
Born term $1/(u - M^2_p)$.  Related, ref.~\cite{korchin} also
details how the Bethe-Heitler process  can be exploited to
extract the TVCS amplitude by measuring the $e^+e^-$ asymmetry.

Lastly, we have performed TVCS calculations for $n(\gamma,\gamma_v)n$
and find similar results, now involving the neutron
form factors. Hence by measuring $d(\gamma,\gamma_v)d$, complimentary
neutron strangeness information can  also be obtained.       
 
\newpage 

Summarizing, we have shown that 
VMD provides a good comprehensive description of meson and nucleon
electromagnetic observables and that
both $\phi$ photoproduction and TVCS are sensitive to VMD parameters.
In particular, TVCS allows probing the better known  $\pi, \eta $ and $\mathcal P$
$t$ channel exchanges at low $|t|$ as well as the uncertain $u$ channel
$\phi N$ coupling at high $|t|$.  Further, 
the TVCS process is ideal
for investigating the time-like nucleon form factor which is 
currently not known below $q^2 = 4 M_{N}^2 \sim 3.5 \; GeV^2$. 
Assuming the validity of VMD with $\phi N$
couplings, we predict narrow, dual peak $e^+e^-$ (and also $\mu^+ \mu^-$) resonances
that should be clearly observable. 
These order of magnitude enhancements
represent a novel experimental signature for both  confirming the 
validity of  VDM  for the $\rho$ and 
$\omega$, which is anticipated, and also for quantifying the $\phi N$
coupling governing the proton's hidden strangeness.


This
work was
partially supported by grants DOE
DE-FG02-97ER41048 and
NSF INT-9807009.  Contributions from  Zach Hill
are also appreciated. 



\newpage

\begin{figure}
\psfig{figure=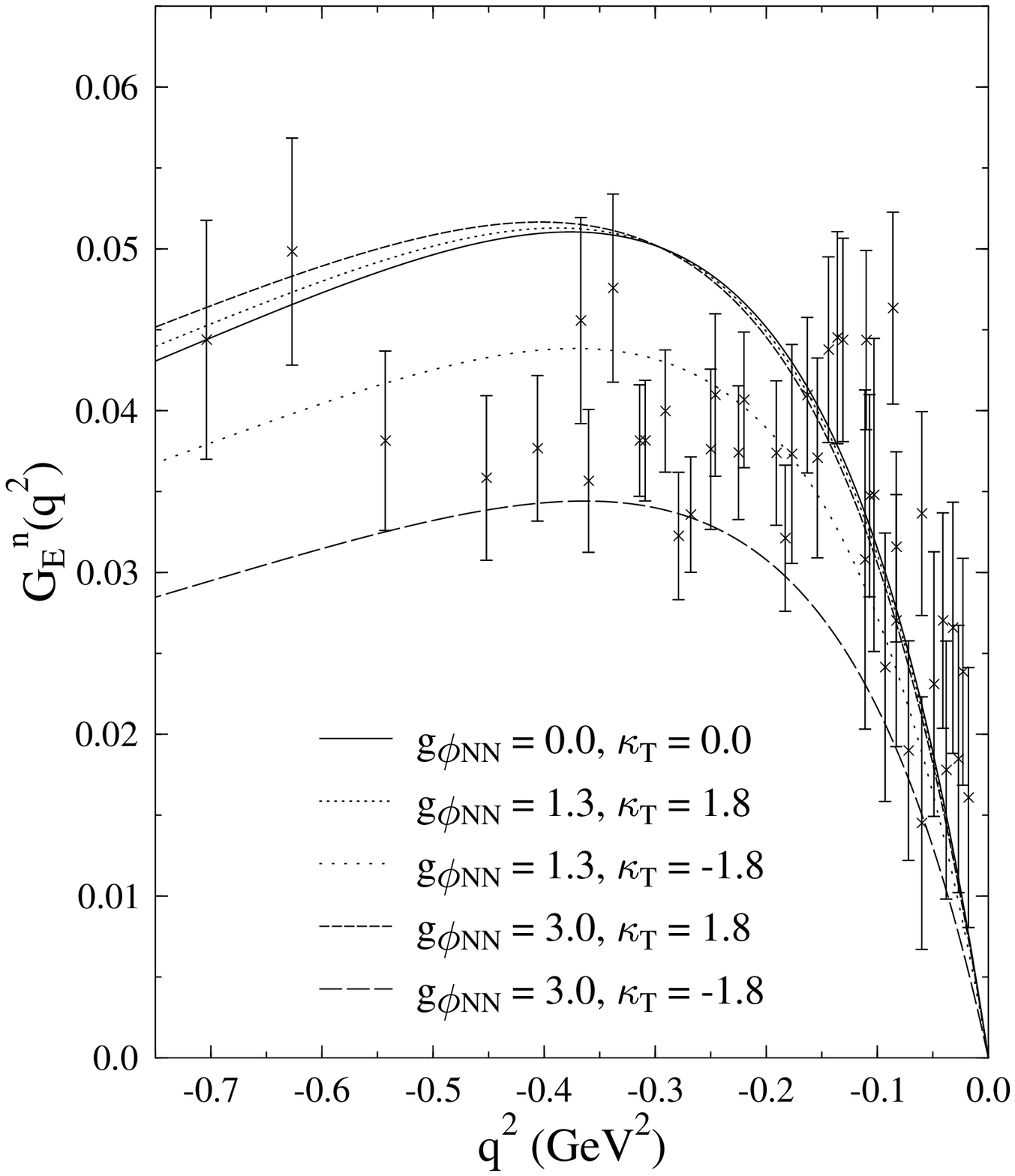,width=5.1in,height=6.6in}
\caption{Data and VMD for the neutron space-like electric form factor.
The curves reflect sensitivity to $\phi N $ coupling.}
\end{figure}

\begin{figure}
\psfig{figure=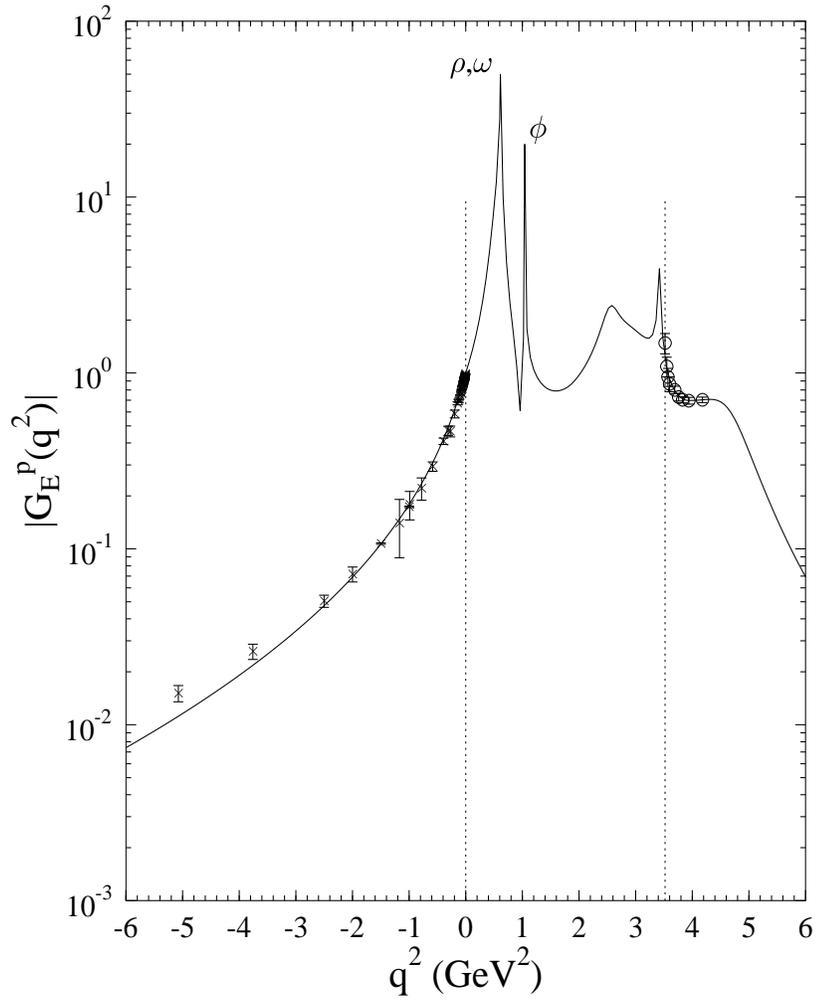,width=5.1in,height=6.6in}
\caption{Data and VMD (absolute value) for the proton electric form factor.
Note the resonant peaks in the unmeasured time-like vector meson region.}
\end{figure}

\begin{figure}
\psfig{figure=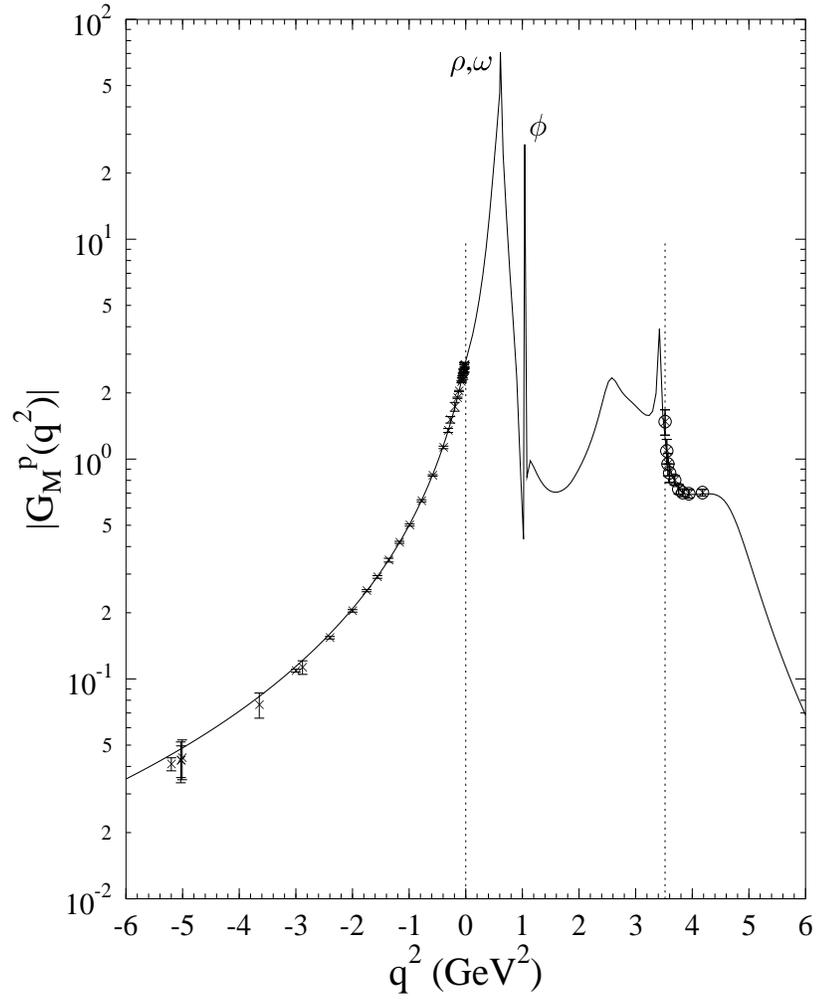,width=5.1in,height=6.6in}
\caption{Data and VMD (absolute value) for the proton magnetic form factor.}
\end{figure}


\begin{figure}
\psfig{figure=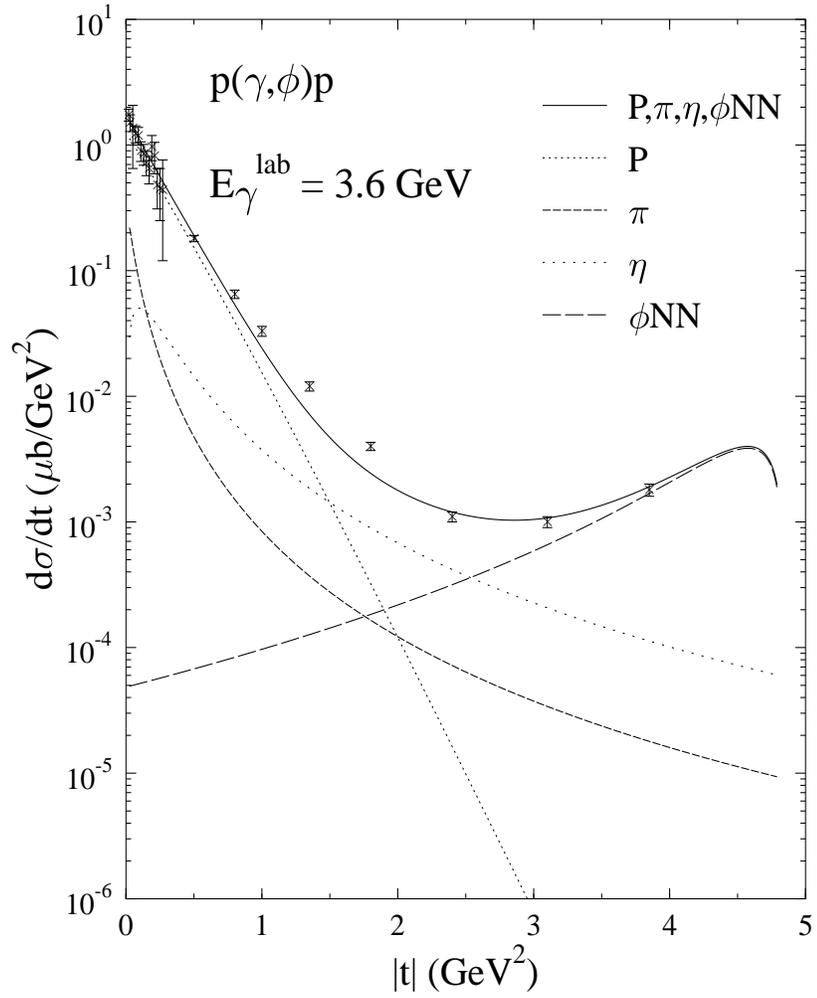,width=5.1in,height=6.6in}
\caption{Theory and data for $\phi$ photoproduction, $p(\gamma,\phi)p$.}
\end{figure}


\begin{figure}
\psfig{figure=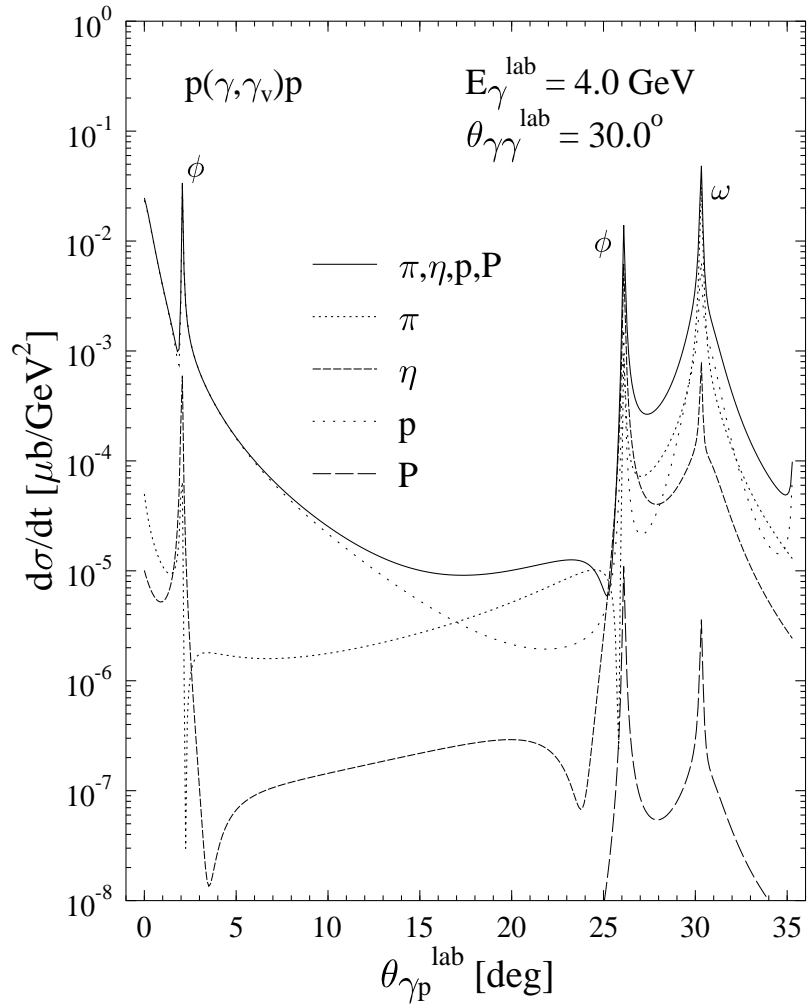,width=5.1in,height=6.6in}
\caption{VMD prediction for TVCS, $p(\gamma,\gamma_v)p$, showing
dual peak resonances.  The smaller angle $\phi$ peak is from $\phi N$ coupling
and quantifies the proton's strangeness.}
\end{figure}


\begin{figure}
\psfig{figure=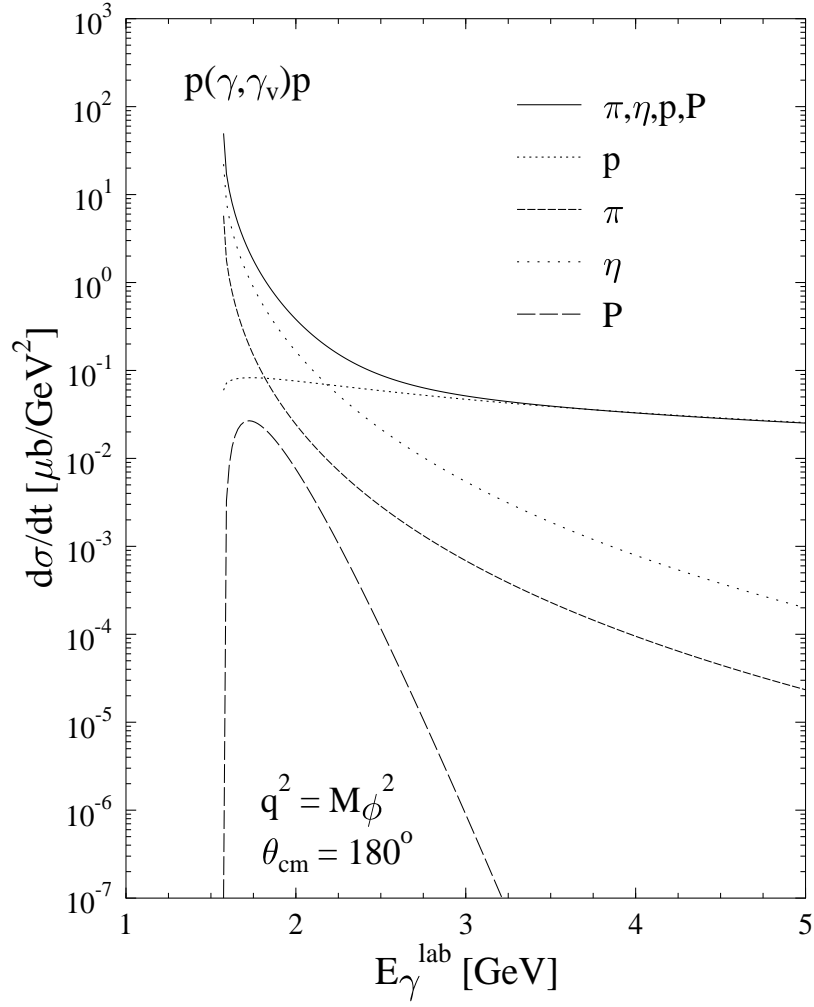,width=5.1in,height=6.6in}
\vspace{1cm}
\caption{VMD prediction for $p(\gamma,\gamma_v)p$ versus lab energy.
The higher energy cross section is dominated by $\phi N$ coupling
(dense dotted line).}
\end{figure}

\end{document}